\documentclass[a4paper, amsfonts, amssymb, amsmath, reprint, showkeys, nofootinbib, twoside, superscriptaddress, longbibliography]{revtex4-1}
\usepackage[english]{babel}
\usepackage[utf8]{inputenc}
\usepackage[colorinlistoftodos, color=green!40, prependcaption]{todonotes}
\usepackage{amsthm}
\usepackage{mathtools}
\usepackage{physics}
\usepackage{xcolor}
\usepackage{graphicx}
\usepackage[left=23mm,right=13mm,top=35mm,columnsep=15pt]{geometry} 
\usepackage{adjustbox}
\usepackage{placeins}
\usepackage[T1]{fontenc}
\usepackage{lipsum}
\usepackage{csquotes}

\newcommand{\fig}[1]{Fig.~\ref{#1}}

\usepackage[acronym,nonumberlist]{glossaries}
\newacronym{NP}{NP}{nanoparticle}
\newacronym{GAP}{GAP}{Gaussian approximation potential}
\newacronym{fcc}{FCC}{face-centered cubic}
\newacronym{hcp}{HCP}{hexagonal closed-packed}
\newacronym{bcc}{BCC}{body-centered cubic}
\newacronym{sc}{SC}{simple cubic}
\newacronym{NS}{NS}{nested sampling}
\newacronym{EAM}{EAM}{embedded-atom method}
\newacronym{DFT}{DFT}{density-functional theory}
\newacronym{MLP}{MLP}{machine learning potential}
\newacronym{PES}{PES}{potential energy surface}
\newacronym{HER}{HER}{the hydrogen evolution reaction}
\newacronym{ML}{ML}{machine learning}

\begin{document}
\title{A general-purpose machine learning Pt interatomic potential for an accurate description of bulk, surfaces and nanoparticles}

\author{Jan Kloppenburg}
    \email[Correspondence email address: ]{jan.kloppenburg@aalto.fi}
    \affiliation{Department of Electrical Engineering and Automation, Aalto University, 02150 Espoo, Finland}
    \affiliation{Department of Chemistry and Materials Science, Aalto University, 02150 Espoo, Finland}
\author{Livia B. P\'artay}
    \affiliation{Department of Chemistry, University of Warwick, Coventry CV4 7AL, UK}
\author{Hannes J\'onsson}
    \affiliation{Faculty of Physical Sciences, University of Iceland,
    VR-III, 107 Reykjav\'ik, Iceland}
    \affiliation{Department of Applied Physics, Aalto University, FI-00076 Espoo, Finland}
\author{Miguel A. Caro}
    \email{mcaroba@gmail.com}
    \affiliation{Department of Electrical Engineering and Automation, Aalto University, 02150 Espoo, Finland}
    \affiliation{Department of Chemistry and Materials Science, Aalto University, 02150 Espoo, Finland}

\date{\today} 

\begin{abstract}
A Gaussian approximation machine learning interatomic potential for platinum is presented. 
It  has been trained on DFT data computed for bulk, surfaces and nanostructured
platinum, in particular nanoparticles. Across the range of tested properties, which include bulk elasticity,
surface energetics and nanoparticle stability, this potential
shows excellent transferability and agreement with DFT, providing state-of-the-art accuracy
at low computational cost. We showcase the possibilities for modeling of Pt systems enabled by this
potential with two examples: the pressure-temperature phase diagram of Pt calculated using
nested sampling and a study of the spontaneous crystallization of a large Pt nanoparticle based on
 classical dynamics simulations over several nanoseconds.
\end{abstract}


\maketitle

\section{Introduction}\label{sec:outline}\label{sec:introduction}

Platinum belongs to the noble metal family and is  
often used in
expensive jewelry. But the wider importance of Pt for the global economy stems
from its countless industrial uses, even in elemental crystalline form.
Platinum is commonly used as a catalyst for many chemical reactions. For instance,
Pt is the best known catalyst for \gls{HER}, where it
shows an extremely small overpotential~\cite{solla_2008,garcia_2010}. Pt is also
one of the few catalysts that can withstand the highly oxidizing environments of the
oxygen reduction reaction (ORR) and oxygen evolution reaction (OER)~\cite{norskov_2004,stephens_2012}.
At the same time, Pt is scarce in
the Earth's crust, and its supply for industrial applications is severely limited by
cost and availability. Still, for some applications, the use of Pt can be so advantageous
compared to the next-best option, that it remains in wide use. To reduce the amount of
raw Pt that is needed for a given application, Pt thin films or
\glspl{NP} can be used instead of the bulk material.
The catalytic properties of Pt films are strongly influenced by crystallographic
surface orientation~\cite{garcia_2010}; for \glspl{NP}, size and shape are the
parameters that determine
these properties~\cite{wang_2008,sanchez_2010,Skulason_2014}. Understanding the atomic-scale
structure of such systems is,
therefore, critical for understanding the catalytic properties.

In this article, we introduce and validate a general-purpose
\gls{ML}-based \gls{GAP}~\cite{bartok_2010,bartok_2015} for elemental Pt. This potential
offers similar accuracy as \gls{DFT} for a small fraction of
the computational cost. Our potential shows extremely good transferability, accurately
predicting the interatomic interactions in Pt from bulk to surface through
\glspl{NP}. The paper is organized as follows. We first discuss the \gls{GAP} theoretical
framework and the generation of training data. We then benchmark our potential against the
prediction of the basic material properties of bulk, surface and \gls{NP} platinum. Finally,
we use the \gls{GAP} to compute the pressure-temperature phase diagram of Pt using the
\gls{NS} method and to study the nucleation of \gls{fcc} Pt during the solidification of a
large Pt \gls{NP}.

\section{Methods} \label{sec:develop}\label{sec:methods}

\subsection{Gaussian approximation potentials}

Gaussian approximation potentials use kernel-based \gls{ML} techniques to regress the
\gls{PES} of an atomistic system. Provided atomic data is available
(typically energies, forces and virials), usually computed at the \gls{DFT} level of theory,
a \gls{GAP} can be \textit{trained} on that data, from which it \textit{learns}. A \gls{GAP}
prediction is made by comparing the atomic structure for which we seek the
prediction to a set of structures in the database. Each of these comparisons yields a
\textit{kernel}, or measure of similarity, which is bounded between 0 (the two
structures are nothing alike) and 1 (the structures are identical). Different
\textit{descriptors}, and combinations thereof, of the atomic structure can be
used to describe the atomic environments. For instance, in this work we use a
combination of two-body (2b) and many-body (mb)
\texttt{soap\_turbo} descriptors~\cite{bartok_2013,caro_2019}.
The actual prediction for the local atomic energy of atom $i$ is expressed as:
\begin{align}
\bar{\epsilon}_i = & e_0 + (\delta^\text{(2b)})^2 \sum_s
\alpha_s^\text{(2b)} k^\text{(2b)} (i, s)
\nonumber \\
& + (\delta^\text{(mb)})^2 \sum_s \alpha_s^\text{(mb)} k^\text{(mb)} (i, s),
\label{eq:gap}
\end{align}
where $k(i,s)$ is the kernel between the atomic environment of $i$ and the
different atomic environments $s$ in the \textit{sparse set} (a subset of structures
in the training database), the $\alpha_s$ are
fitting coefficients obtained during training, $e_0$ is a constant energy per
atom and $\delta$ gives the energy scale of the model. Forces can be obtained
analytically from the derivatives of Eq.~(\ref{eq:gap}). More details about the \gls{GAP}
framework and many-body descriptors are given in Refs.~\cite{bartok_2010,
bartok_2013,bartok_2015,caro_2019}.

We generate training data at the \gls{DFT} level using the PBE functional approximation~\cite{perdew_1996}
and will denote it as PBE-DFT
from now on. We use a highly converged
plane-wave basis set with a 520~eV cutoff and an adaptive reciprocal-space integration
mesh such that the number of \textbf{k} points is given by
$N_\textbf{k} = 1000 / N_\text{atoms}$. The VASP software~\cite{kresse_1996,kresse_1999,bloechl_1994}
is used with input options given in the appendix. The composition and generation
of the database are discussed in Sec.~\ref{sec:database}. The training and
validation of the potential are done with the QUIP/GAP codes~\cite{ref_quip}.
Structure manipulation and database
sorting are done with ASE~\cite{larsen_2017}. Molecular dynamics (MD) simulations
are carried out using LAMMPS~\cite{plimpton_1995,ref_lammps} and
TurboGAP~\cite{ref_turbogap}.

\subsection{Database generation and accuracy tests}\label{sec:database}

We want to create a robust Pt GAP that can be used safely in exploratory
work, e.g., to assess the stability of Pt \glspl{NP} derived computationally.
To this end,
the GAP needs to be simultaneously accurate \textit{and} transferable. Within
a data-driven approach, it is important to note that prior knowledge of physics and
chemistry is not embedded in the form of the potential. That is, a GAP does not ``know''
about the Schr\"odinger equation -- it only knows about data it has seen during training.
Therefore, the training set must be carefully crafted to contain all the relevant
configurations. This includes (meta)stable structures, but also, perhaps counterintuitively,
high-energy structures. High-energy structures must be present in the database so that
the GAP learns that they are, in fact, of high energy, otherwise the GAP could spuriously
predict previously unseen unstable structures to be low in energy.

It is also useful to realize that high-energy observables can be learned with less
accuracy than low-energy ones, because low-energy structures contribute much more to
the partition function of the system at the temperatures of interest,
and thus to the derived thermodynamic properties.
This leads to an efficient database construction strategy where
a few disordered structures, such as high-temperature liquid or dimers at close range, are
added to sample configuration space sparsely but comprehensively. Further to these, many
configurations close to the known stable structures, like close-packed crystal lattices
and surfaces thereof, are added by ``rattling'' the atoms around equilibrium and applying
small amounts of strain to the periodic cells. This, in turn, begs the question: what about
the \textit{unknown} stable structures?

To improve a GAP in a yet unknown region of configuration space,
a successful strategy is iterative
training~\cite{deringer_2017}. In iterative training one trains several versions of the
GAP, and each time uses the newest GAP to predict stable structures. The energy values and atomic forces
for those structures are then computed with PBE-DFT and fed to the next version of the GAP, which
will learn from its predecessor's successes and, especially, failures. This iterative procedure
progressively refines the GAP's accuracy in the region of configuration space where the target
structures (e.g., \glspl{NP}) reside. The advantage is that the computationally demanding
procedure, the structure generation, which might require thousands or millions of 
energy and force evaluations, is performed with the GAP, inexpensively. The PBE-DFT calculations
are only carried out for the final structures or, in some cases, a small subset of the
structures selected along the path followed in configuration space to generate the final ones.

\begin{figure}
    \centering
    \includegraphics[width=\columnwidth]{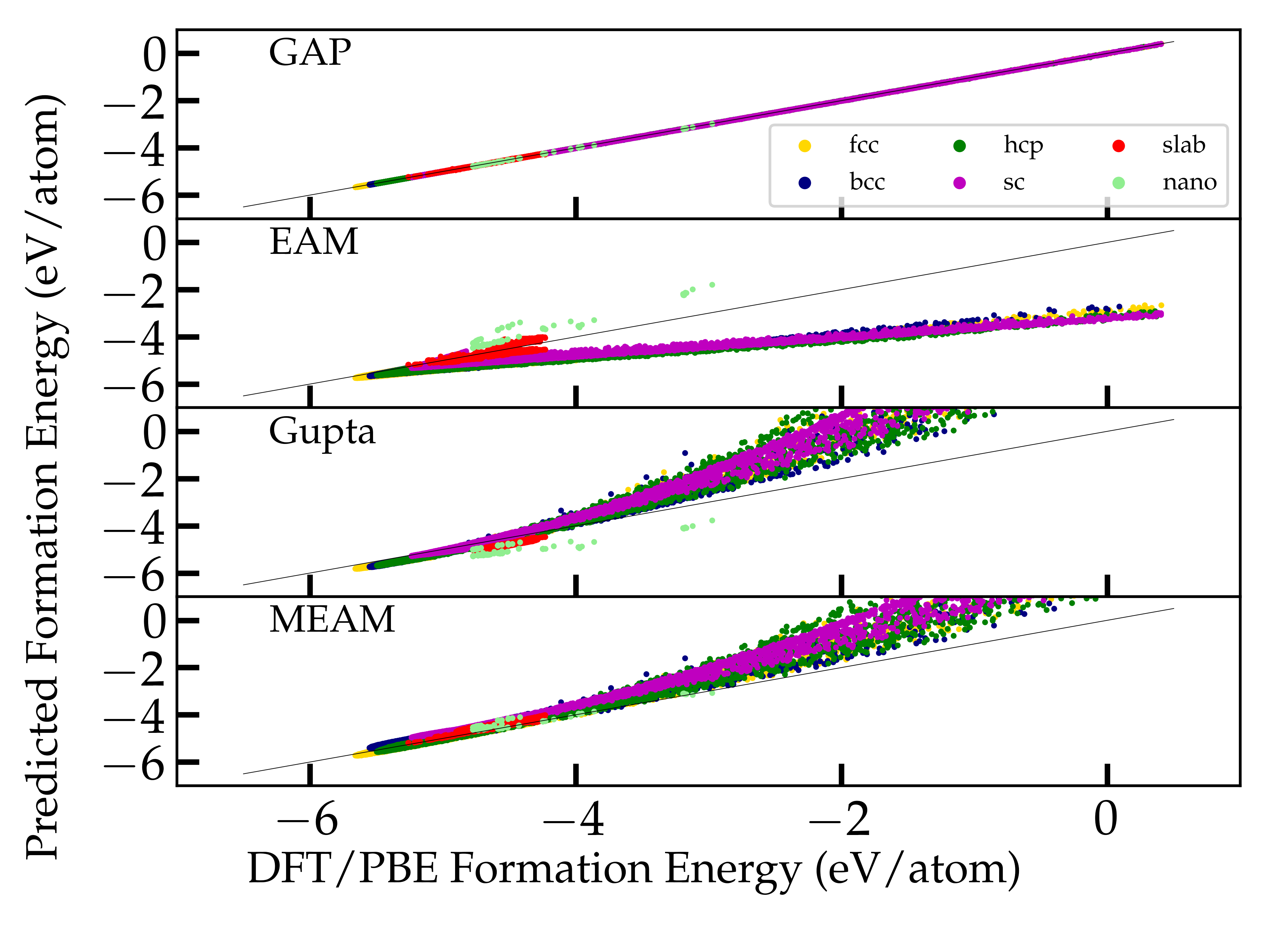}
    \caption{Validation of the Pt GAP performed on atomic configurations
    unseen during training. The different configuration types (\gls{fcc}, \gls{hcp}, etc.) are
    indicated with different colors. The results of testing an EAM, Gupta and MEAM
    potential on the same set of structures are given for reference. Formation energies
    are computed by $E_\text{f}=E_\text{pred} - n\mu$, where $n$ gives the number of atoms in the
    structure and $\mu$ is the reference energy per atom for the given potential (e.g., the energy of
    an isolated Pt atom). The GAP errors are: maximum energy error:
    0.043~eV/atom; energy RMSE: 1.6 meV/atom; and force RMSE: 0.111~eV/\AA{}.
    The EAM, Gupta and MEAM errors are, respectively:
    maximum energy error: 3.1, 7.7 and 4.8~eV/atom; energy RMSE: 420, 1457, and 607~meV/atom;
    and force RMSE: 1.36, 1.69 and 1.21~eV/\AA.}
    \label{fig:validate}
\end{figure}

\begin{figure}
    \centering
    \includegraphics[width=\columnwidth]{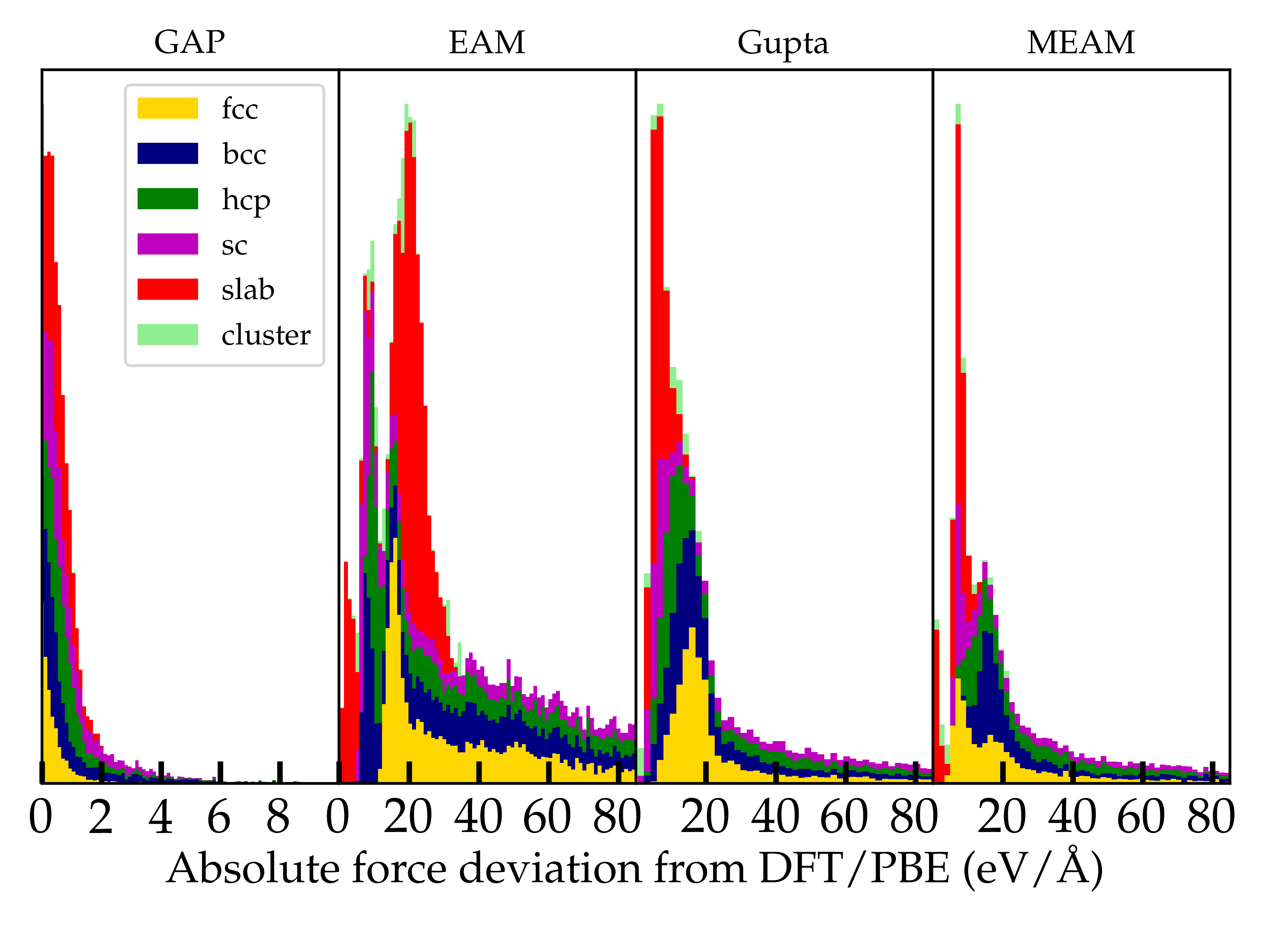}
    \caption{Force-error stacked histograms for our GAP and the tested reference
    potentials. Note that the ranges are one order of magnitude wider for EAM, Gupta and MEAM,
    compared to the GAP. The distributions are color coded according to structure type. The
    corresponding RMSEs are given in the caption of \fig{fig:validate}.}
    \label{fig:force_hist}
\end{figure}

Figure~\ref{fig:validate} shows the most commonly used numerical benchmark for
\glspl{MLP}, i.e., a scatterplot of predicted versus reference energies for
a 20\%/80\% test/training sets split. That is, out of the entire database of
structures, 80\% are used to train the GAP and the 20\% unseen structures are
used to test the potential outside the training set.
The root mean-square error (RMSE) is computed to give a single numerical
score for the overall performance of the potential. Our Pt GAP shows remarkably
low errors in this simple test, with an RMSE of only 1.6~meV/atom.
Application-specific tests of the GAP are presented in the following section,
more indicative of how this potential performs for large-scale and high-throughput
simulations.

In \fig{fig:validate} we also show the predictions of three \gls{EAM}-type
potentials on our test set for comparison: Zhou's
EAM~\cite{ZHOU20014005,PhysRevB.69.144113}, the Gupta potential~\cite{B707136A} and Lee's modified EAM (MEAM)~\cite{lee_2003}.
EAM-type potentials are usually fitted using ground-state
(low-temperature) experimental data.
For instance, Zhou's EAM was fitted to reproduce (quoting the authors) ``basic material properties such
as lattice constants, elastic constants, bulk moduli, vacancy formation energies,
sublimation energies, and heats of solution''~\cite{ZHOU20014005}.
While \glspl{EAM} can satisfactorily reproduce the
energetics of bulk \gls{fcc} near equilibrium, as expected from the composition
of their training database, all other structure
types are modeled significantly less accurately. We will show in Sec.~\ref{sec:nanoparticles}
a further comparison for \glspl{NP}.
In \fig{fig:force_hist} we show a histogram of force errors. Systematic
deviations for the reference potentials (i.e., the error distributions peak above zero) are
apparent, also visible in \fig{fig:validate} for high-energy structures. These high-energy
structures fall outside the scope of EAM-type potentials. However, they become relevant in
low-dimensional systems and, e.g., at high temperature and/or pressure.

Clearly, the improved accuracy of \gls{GAP} comes at
the expense of additional CPU time. For instance, to perform a single-point calculation for
a \gls{NP} with 147 atoms, our \gls{GAP} requires approximately 109~ms
of CPU time whereas an \gls{EAM} calculation only needs 1.2~ms. The \gls{GAP}
is still significantly faster than PBE-DFT (using VASP), for which this calculation requires
of the order of 102 CPU hours (i.e., $\sim 3.5 \times 10^6$ times more expensive
than the \gls{GAP}).

\subsection{Nested sampling}

We use the nested sampling (NS) technique~\cite{skilling_2004,NS_all_review} to evaluate the bulk
macroscopic thermodynamic properties of the new Pt \gls{GAP} model.
NS samples the entire potential energy surface, starting from high-energy
random configurations (representing the gas phase) down to the ground-state structure
through a series of nested energy levels, without requiring any advance knowledge of
the stable  phases~\cite{EfficientSampling,NS_mat_review}.
A unique advantage of NS is that it allows the calculation of the partition function
as a simple post-processing step. 
This gives access to thermodynamic properties, such as the heat capacity--which is
the second derivative of the partition function with respect to temperature--and
hence enables us to identify all the phase transitions of the system. 
In the current work we perform the NS calculations at constant pressure, to compute
the pressure-temperature phase diagram~\cite{pt_phase_dias_ns,ConPresNS,NS_Iron,NS_lithium}. 
Simulations were carried out using the pymatnest program package~\cite{pymatnest},
using LAMMPS to perform the dynamics.


\section{Benchmarks} \label{sec:benchmarks}

Our Pt GAP has been designed with the goal of general applicability in mind. In this section we prove
its transferability across a selection of different applications representative of common use
cases. We test the GAP for basic bulk properties (equation of state, elasticity
and phonons), surface energetics and \gls{NP} formation energies. While avoiding a too detailed
examination of each application, which could merit on their own more focused studies,
these examples showcase the predictive power of the
new GAP. In Sec.~\ref{sec:applications} we describe two, more detailed, applied studies:
the phase diagram and a spontaneous crystal nucleation in nanostructured Pt.

Accuracy tests that are missing from this section are those
relevant to nanoscopic processes in surface diffusion, aggregation, nucleation and, more
generally, rare events involving the description of a transition state while crossing an
energy barrier. The exploration of the region of configuration space corresponding to
these is difficult to automate, because transition states contribute much less to the
partition function than stable states and will not be sampled by MD in significant
proportions. We are currently developing the methodology to automate incorporating
transition-state configurations for GAP training, and will add these to future versions
of our Pt GAP as these developments become available. In the meantime, one should expect
inconsistent prediction of transition-state energies with our GAP and individual calculations
should be benchmarked against DFT before trusting the results.

\subsection{Equation of state, elastic properties and phonons}

\begin{figure}
    \centering
    \includegraphics{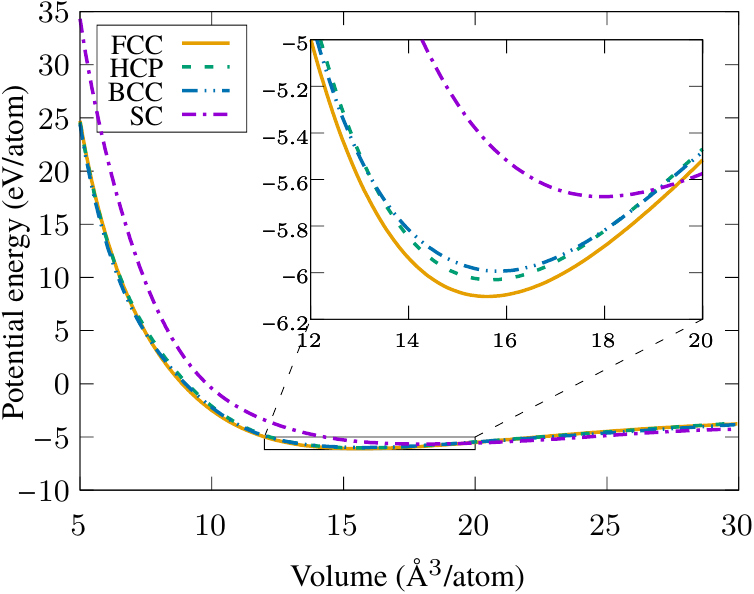}
    \caption{Equation of state for three cubic and one hexagonal Pt crystal phases:
    \acrfull{fcc}, \acrfull{hcp}, \acrfull{bcc}
    and \acrfull{sc}. The inset shows a closeup in the region where the global minimum is
    located.}
    \label{fig:eos}
\end{figure}

The equation-of-state calculation shows the expected minimum for the \gls{fcc} phase from zero
up to very high pressure, with \gls{hcp} about 60~meV/atom above \gls{fcc} and \gls{bcc} slightly above \gls{hcp}.
The \gls{sc} phase is significantly higher in energy than \gls{fcc}, \gls{hcp} and \gls{bcc},
except at  large tensile strain, i.e., at  large (and unrealistic) negative
pressures, where it becomes the stable phase. All phases evolve smoothly as a function of
unit cell volume, as shown in \fig{fig:eos}.

\begin{figure}
    \centering
    \includegraphics[width=\columnwidth]{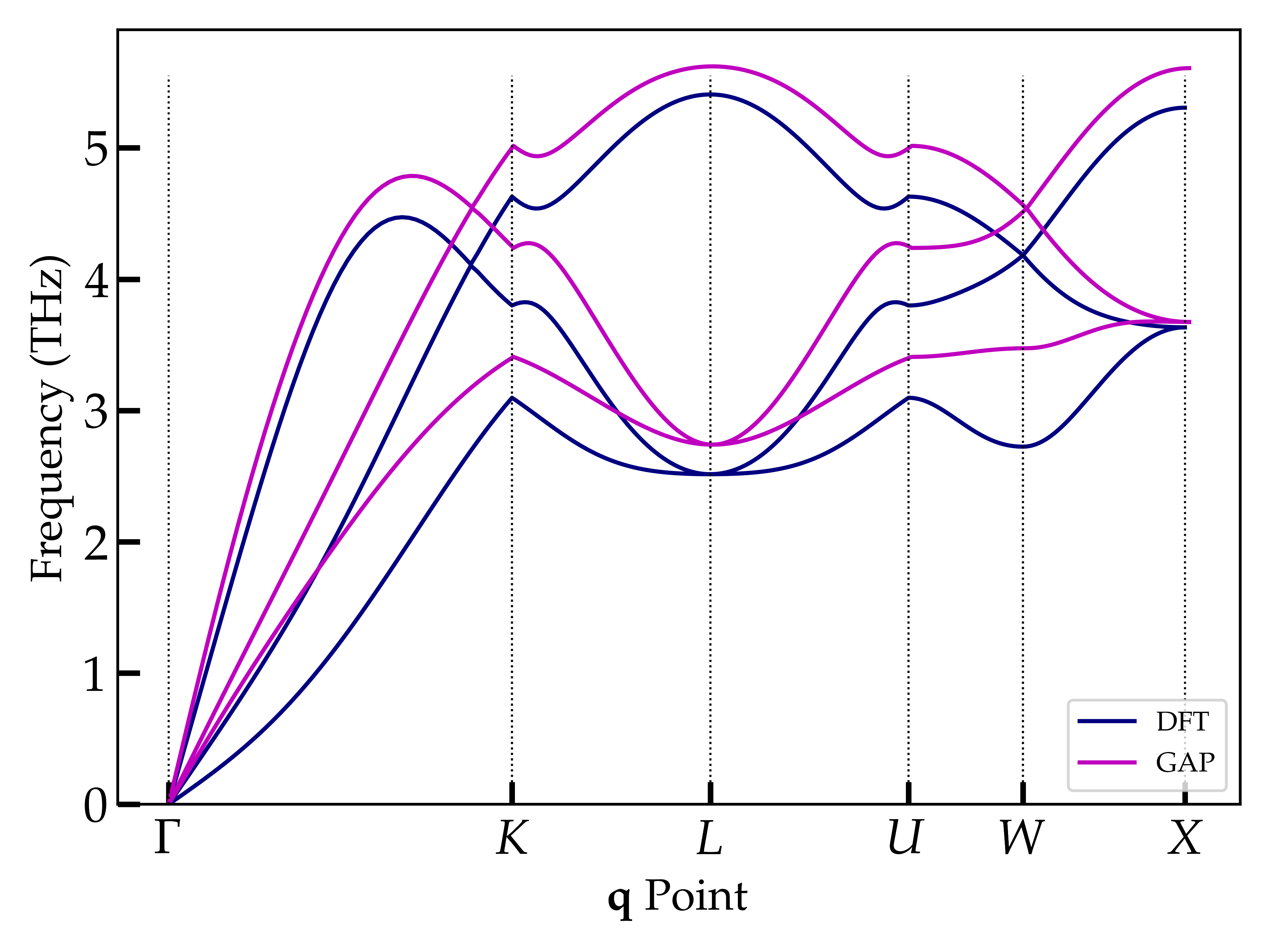}
    \caption{Phonon dispersion as computed by GAP and PBE-DFT with
    phonopy~\cite{phonopy}. The trends are well reproduced in comparison
    to PBE-DFT except for a systematic deviation at the \textit{W} point.}
    \label{fig:phonons}
\end{figure}

Our tests for Pt show that phonons and elastic constants can be learned accurately when
the training data only contains structures created for this specific purpose.
However, the trained potential is then only
able to describe those properties, and will not have general-purpose applicability.
When different structures are added to bring in more general-purpose applicability, the high accuracy on both phonons and
elastic constants is sacrificed. Phonons (\fig{fig:phonons}) are still described reasonably well
as compared to PBE-DFT results as far as the main trends are concerned,
except for a systematic deviation at the \textit{W} point. Table~\ref{tab:elastic} shows the elastic constants
computed with GAP and compared to PBE-DFT, as well as to experiment. Overall, the agreement with
PBE-DFT (the GAP's ``ground truth'') is good, with a systematic deviation of only $+4$\%. This
deviation is smaller than the PBE-DFT error as compared to experiments, highlighting how, in some
cases, the overall accuracy of the GAP is more limited by the intrinsic
accuracy of the reference method (PBE-DFT in our case) than by the accuracy of the fit.
That is, for the specific purpose of calculating elastic constants, the GAP is a better representation
of PBE-DFT than PBE-DFT is of reality.
For reference,
we also provide the elastic constants computed with EAM, which compare favorably to
experiment. They are indeed closer to the experimental values than the PBE-DFT results,
a consequence of the fact that the EAM was fitted to reproduce experimental ground-state
results, as discussed earlier.

\begin{table}[t]
\caption{Comparison of GAP-predicted elastic constants with PBE-DFT as well as experimental values. The
percentage in brackets shows the deviation vs experiment for PBE-DFT and EAM,
and the deviation vs PBE-DFT for GAP.}
\centering
\begin{ruledtabular}
\begin{tabular}{lcccc}
& Exp.~\cite{collard_1992} & PBE-DFT & GAP & EAM \\ \hline
$C_{11}$ (GPa) & 373 & 320 ($-14$\%) & 333 ($+4$\%) & 345 ($-8$\%) \\
$C_{12}$ (GPa) & 242 & 218 ($-10$\%) & 228 ($+5$\%) & 250 ($+3$\%) \\
$C_{44}$ (GPa) & 78 & 77 ($-1$\%) & 80 ($+4$\%) & 76 ($-3$\%)
\end{tabular}
\label{tab:elastic}
\end{ruledtabular}
\end{table}

\subsection{Surfaces}\label{sec:surfaces}

Platinum is a material widely used in interfacial (electro)catalysis, and thus it is
important to ensure that an interatomic potential for Pt can accurately reproduce surface
formation energies. The three surfaces most commonly studied are those defined by the (111),
(100) and (110) crystallographic \gls{fcc} planes~\cite{garcia_2010}.
The (111) surface is the most stable one and the one most often used in electrocatalysis,
e.g., for hydrogen production, due to the low overpotential it exhibits for \gls{HER}~\cite{solla_2008}.

A comprehensive study of surface
energetics for arbitrary Miller indices $(hkl)$ becomes prohibitive for DFT, due to the large number of atoms in the unit cell for large indices. For example, a 7-atom thick
Pt slab with $(10\, 1 \, 0)$ indices already contains 280 atoms in the \textit{primitive}
unit cell. With our Pt GAP, studying these surfaces with small tilt angles becomes possible.
We therefore calculated the surface formation energies,
with bulk \gls{fcc} Pt as reference, for all the symmetry-inequivalent Miller planes that
can be constructed in Pt when letting each index run up to 10. To ensure that
reconstruction effects beyond the primitive unit cell are considered, we ran the
calculation for the primitive unit cell generated with ASE~\cite{larsen_2017} as implemented
by Buus, Howalt and Bligaard,
its Niggli equivalent cell~\cite{niggli_1929,krivy_1976,grosse_2004},
as well as $2 \times 2$ supercells built thereof. We included
small random initial displacements of the atoms to avoid biasing the geometry optimization due
to high-symmetry starting configurations. Altogether, six calculations were carried out
for each set of Miller indices and the obtained surface formation energies per atom
were always the same (except for negligible numerical differences). This indicates that
simple relaxation of the atomic positions takes place as the surfaces are created 
and that the surfaces have the same periodicity as the primitive unit cell.

\begin{figure}
    \centering
    \includegraphics{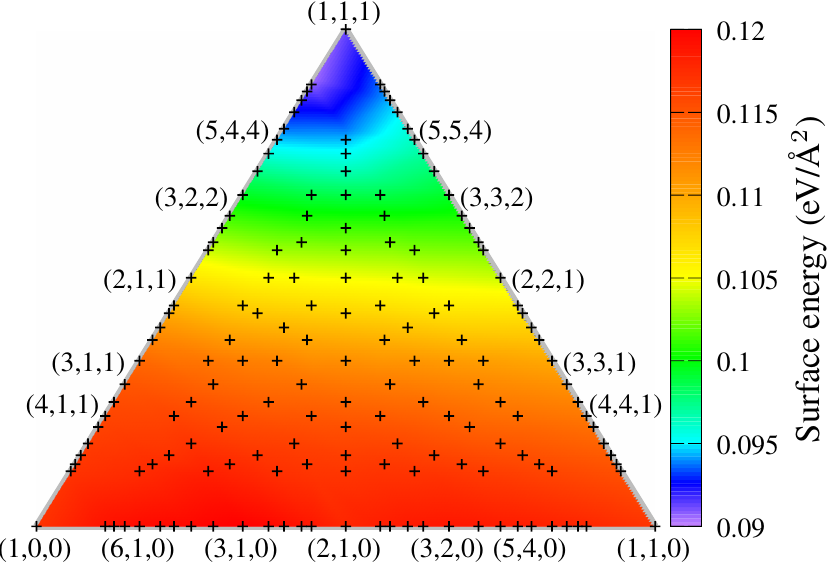}
    
    \vspace{0.5cm}
    
    \includegraphics[width=\columnwidth]{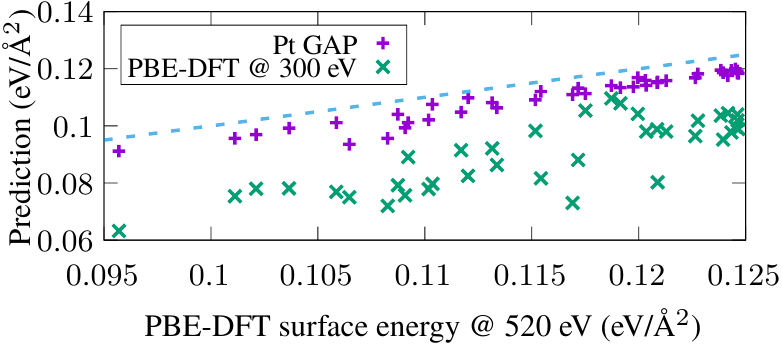}
    \caption{(Top) Surface energies computed with the Pt GAP for a range of crystal
    orientations resulting from tilting the faces between the (111), (100) and (110)
    directions. Each cross represents an actual data point, with selected Miller indices
    indicated, and the color map is drawn by interpolating
    between those. (Bottom) Comparison between the Pt GAP, a standard-quality PBE-DFT calculation
    (with VASP defaults and a 300~eV plane-wave cutoff), and our reference VASP PBE-DFT calculations
    (which use a larger cutoff of 520~eV).}
    \label{fig:surface_triangle}
\end{figure}

Figure~\ref{fig:surface_triangle} (top panel) shows the surface formation energies for varying
Miller indices within the triangle enclosed by the (111), (110) and (100) planes as
end points. The values predicted for those planes, 0.091, 0.117 and 0.117~eV/\AA$^2$,
respectively, are in good agreement with our reference PBE-DFT values (0.096, 0.123 and
0.120 eV/\AA$^2$, respectively) and with recent values
from the literature~\cite{tran_2016}.
The GAP predicts smooth transitions as the cleaved (and relaxed) crystal facet is tilted
between the most common facets. The bottom panel of the figure shows the surprising
result that our Pt GAP is more capable of reproducing the PBE-DFT surface energies provided
by a ``high-quality'' PBE-DFT calculation (computed with the same VASP settings as those
reported in Sec.~\ref{sec:methods}) than another PBE-DFT calculation with ``standard''
settings. The GAP tends to slightly underestimate the surface formation energies (by
about 5\%) but the trends, i.e., the relative formation energies, are accurately captured.

\subsection{GAP accuracy for nanoparticle modeling}\label{sec:nanoparticles}

We have generated a large database of Pt \glspl{NP} for this work.
This database is divided into the two following subsets,
NP-DB01 and NP-DB02.
NP-DB01 contains 8000 \glspl{NP} generated between $N_\text{atoms} = 10$ and $N_\text{atoms} = 349$
using an annealing-quenching-relaxation procedure, starting from a highly disordered precursor,
where the annealing and quenching steps take 20~ps each and the annealing happens at 1500~K; we
call this a ``cooking'' protocol. NP-DB02 contains 3400 \glspl{NP} between
$N_\text{atoms} = 10$ and $N_\text{atoms} = 349$ (10 for each size) where the annealing step of
the cooking protocol takes place at the optimal crystallization temperature of 1150~K
(see Sec.~\ref{sec:spontaneous_nucleation}) but
otherwise generated in the same way as NP-DB01.
This database is freely available from the Zenodo repository~\cite{kloppenburg_2022d}
and will be extended in
subsequent work, in particular with larger \glspl{NP} beyond $N_\text{atoms} = 349$.

\begin{figure}[t]
    \centering
    \includegraphics[width=\columnwidth]{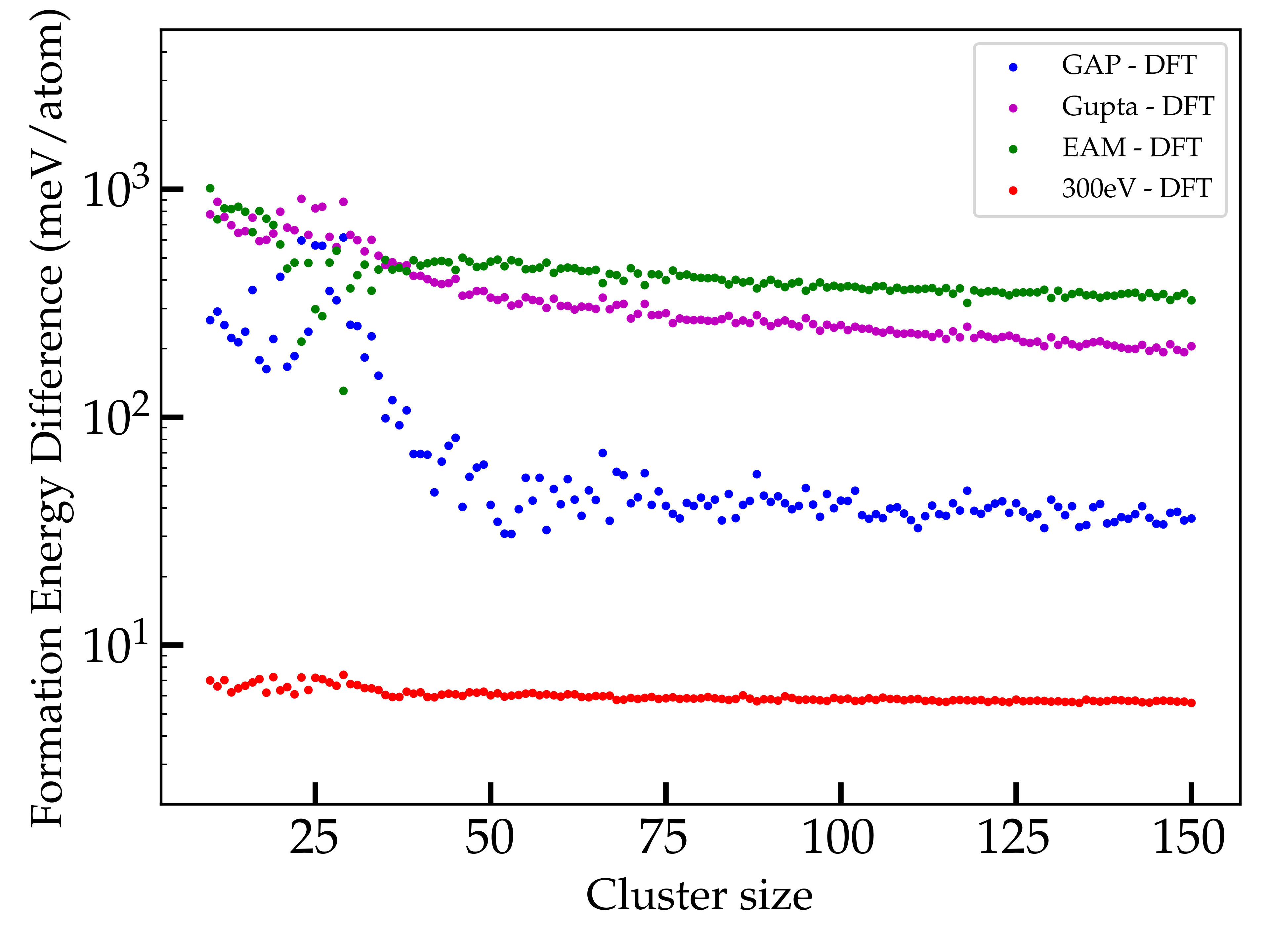}
    \caption{Formation energies for a selection of annealed \glspl{NP} computed with different
    potentials and standard-quality PBE-DFT versus a benchmark-quality PBE-DFT calculation.}
    \label{fig:nanoparticles}
\end{figure}

To assess the ability of our GAP to accurately model Pt \glspl{NP} and to compare it to previously
available, commonly used, force fields for Pt modeling, we selected \glspl{NP} from NP-DB01
up to $N_\text{atoms} = 150$. The energies were computed with our GAP, standard-quality PBE-DFT
(300~eV plane-wave cutoff), the Gupta potential~\cite{B707136A}, and the \gls{EAM}
potential~\cite{ZHOU20014005,PhysRevB.69.144113}.
We compare all these numbers to a benchmark-quality PBE-DFT calculation (520~eV
plane-wave cutoff). The results of this comparison are shown in \fig{fig:nanoparticles}. Clearly,
the GAP outperforms the other force fields with errors ($\sim 40$~meV/atom) one order of
magnitude smaller than Gupta ($\sim 400$~meV/atom) and \gls{EAM} ($\sim 500$~meV/atom) around and
above 50-atom \glspl{NP}. The GAP errors for these
\glspl{NP} are about 5 times larger than those obtained from standard-quality PBE-DFT.
For very small \glspl{NP} (< 50 atoms) the GAP results are still better than for
the other force fields but the errors are significantly higher than for larger \glspl{NP}.
Since the atomic motifs in small \glspl{NP} look significantly different from those of bulk and surfaces,
it is not surprising that the errors are larger.

\begin{figure}[t]
    \centering
    \includegraphics[width=\columnwidth]{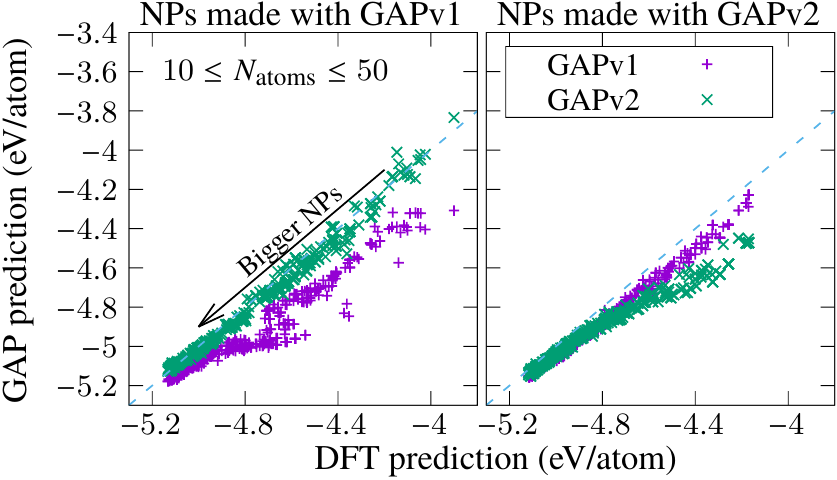}
    \caption{(Left) Predicted GAPv1 and GAPv2 energies computed on \glspl{NP} generated with
    GAPv1, versus the corresponding \gls{DFT} values, for \glspl{NP} in the size range from 10 to
    50 atoms. High energy-per-atom values correspond to smaller \glspl{NP} whereas low values correspond
    to larger \glspl{NP}. (Right) Same test as on the left panel but performed on \glspl{NP} generated
    with GAPv2. The followed \gls{NP} generation method in both cases is the ``cooking'' protocol
    reported in the text.}
    \label{fig:iterative_nps}
\end{figure}

The accuracy of \gls{GAP} can be enhanced
specifically for \glspl{NP} by iteratively training the potential for that purpose.
That is, we can improve the accuracy of
the GAP in the future by adding (some of) these \glspl{NP} to the training set and training a new
version of the potential, as exemplified in \fig{fig:iterative_nps}. In that figure we observe the
performance of two versions of the Pt \gls{GAP}. The first one, GAPv1, is initially used to make two sets
of small \glspl{NP}, with $N_\text{atoms} \le 50$. One of the sets is used to retrain the
\gls{GAP}, giving GAPv2, and the second set is used to test the predictions of both versions versus
PBE-DFT. The results are shown on \fig{fig:iterative_nps} (left) where GAPv1 is shown to predict
too low (i.e., too stable) energies for the smallest \glspl{NP} in the test set
($ N_\text{atoms} \lesssim 40$)
whereas GAPv2 correctly orders all of the \glspl{NP} generated with GAPv1. On the right-hand side of the
figure we show the energy predictions of GAPv1 and GAPv2 for \glspl{NP} that were generated with
GAPv2. There are two features of the \gls{GAP} accuracy refinement provided by iterative
training which are apparent from this right-hand panel. On the one hand, as expected, GAPv2 produces
``better'' \glspl{NP} than GAPv1, in the sense that they are lower in energy when looking at the
PBE-DFT energy prediction (i.e., the datapoints are shifted horizontally to the left, compared
to the left-hand panel), and there is less
data scatter. On the other hand, counterintuitively, the GAPv1 predictions for these GAPv2-generated
\glspl{NP} are in better agreement with PBE-DFT than the GAPv2 predictions. While unexpected, this
is a typical result for early iterations in \gls{GAP} iterative training: a given iteration of the potential,
used in an application-specific simulation, will favor structures which populate artificially low
regions of the \gls{PES}. As new iterations of the potential add these low-energy structures
to the database, the \gls{PES} is refined and the \gls{GAP} ``unlearns'' the spurious minima and
the scatterplot converges towards optimal agreement with \gls{DFT}.

The structure generation strategy that we followed here to augment the GAPv1 database
is as follows. We first generate all
the regular FCC tetrahedra that can be constructed below 50 atoms, which correspond to 4, 10, 20 and 35 atoms.
We then start an MD simulation from the ideal (relaxed) structure, quickly (10~ps) heat up to 3000~K and quickly
(another 10~ps) quench down to 100~K. From this MD trajectory, we sample 11 equidistant (in time) snapshots, which ensures we
incorporate a wide diversity of small nanoclusters, including some that are high in energy: regular (crystal-like),
thermally disordered and quenched structures are added to the training database.

Generally, as new training configurations are generated, we can retrain and refine the accuracy of our
\gls{GAP}. For reference, we provide in the repository~\cite{kloppenburg_2022c} two versions
of the \gls{GAP}: the one used for most of the
simulations presented in this article (v1) and the one that contains a small amount of
\gls{NP}-specific iterative training (v2). Any future version of the GAP will be added to this repository together with a note on any further additions to the database,
compared to the configurations reported here, with all published versions remaining publicly
available. This will ensure that the user base of the potential has easy access to the most accurate
(and most recent) Pt \gls{GAP} while enabling reproducibility of the results produced with all
earlier versions. Upcoming work from our group will focus on a detailed study of small Pt \gls{NP}
formation and stability, and we expect to update this repository with a \gls{NP}-optimized version
of the \gls{GAP} in the near future.

For the sake of clarity, we emphasize here that GAPv1 was used to generate all the
results in this paper except for those labeled as GAPv2 results in this subsection. In addition,
we note that the linked repository~\cite{kloppenburg_2022c} allows to browse the full history of
GAP versions, even though the latest version is shown by default. Both v1 and v2 can be retrieved
from the repository and are listed under ``Versions''.


\section{Applications}\label{sec:applications}

\subsection{Pressure-temperature phase diagram}

The NS calculations were performed as presented in Ref~\cite{ConPresNS}.
The simulations were run at constant pressure in the range of $p=0.07-50\mathrm{~GPa}$,
using a simulation cell of variable shape and size, containing 24 atoms. 
We used 1000 walkers and performed 440 steps (8:1:2:2 ratio of total-energy
Hamiltonian Monte Carlo, volume, cell shear and cell stretch steps) to generate the
new configurations during the NS iterations. These parameters ensure convergence of
the melting transition within $\pm40$~K.  
The use of small systems will inevitably cause some finite-size effects, for example an
underestimation of the boiling curve and an overestimation of the melting line
as compared to the macroscopic value~\cite{pt_phase_dias_ns}.
In order to estimate this error, we repeated the simulations with 48 atoms at
$p=1~\mathrm{GPa}$, and obtained 2.8\% lower melting temperature as compared to the 24-atom
calculation. 

\begin{figure}[t]
    \centering
    \includegraphics[width=8.5cm]{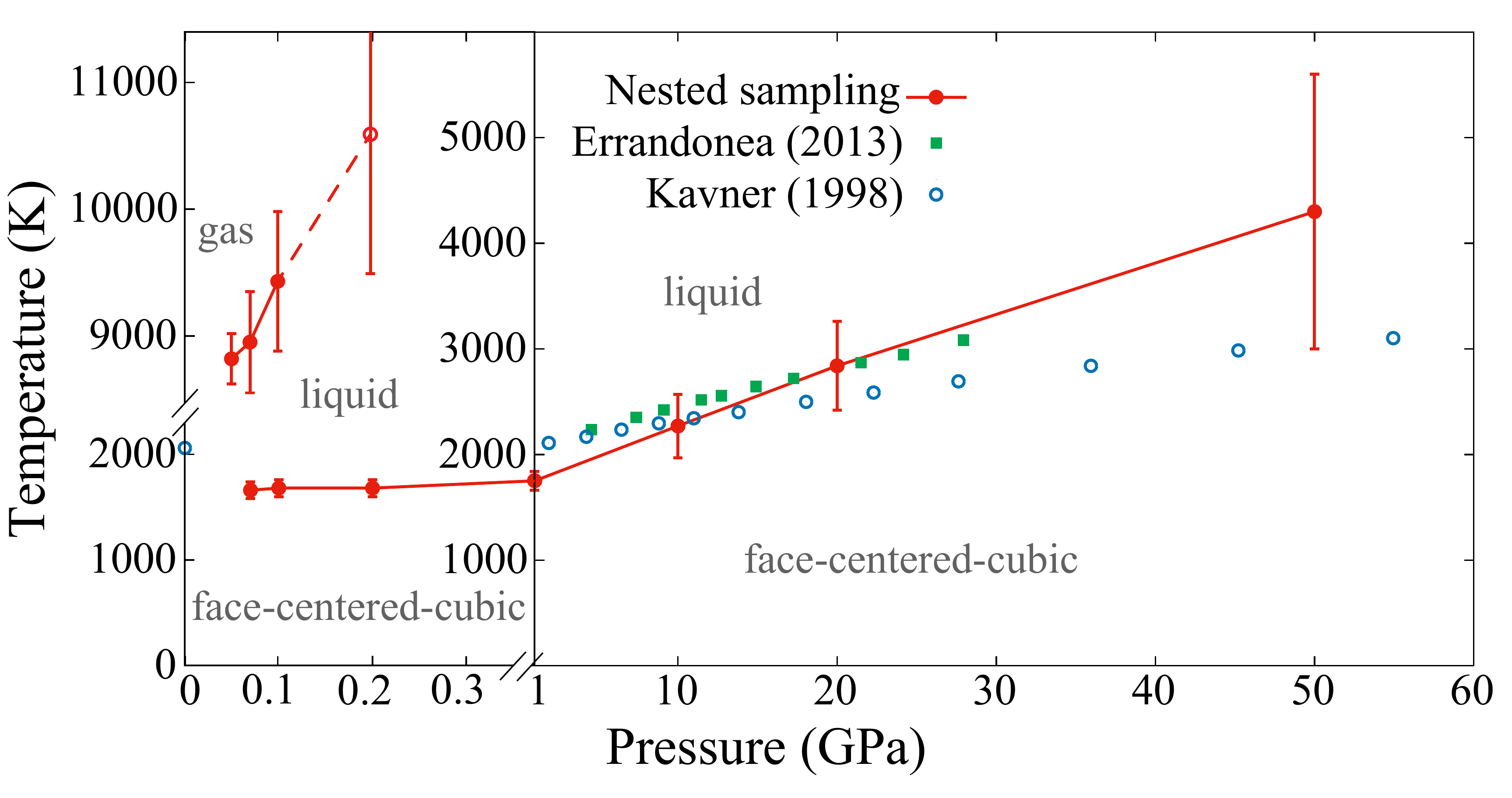}
    \caption{Pressure-temperature phase diagram calculated by nested sampling (red lines and symbols). Error bars represent the full widths at half maximum of the heat capacity curves. Green and blue symbols show experimental melting temperatures taken from Refs.~\cite{Pt_melt_Errandonea} and \cite{Pt_melt_Kavner}, respectively.}
    \label{fig:pt_diagram}
\end{figure}

Figure~\ref{fig:pt_diagram} shows the pressure-temperature phase diagram.
At low pressure, we observe a heat capacity peak at high temperature corresponding
to the boiling curve and its extension to the supercritical region, the Widom line,
marked by a shallower and broader peak (shown by dashed red line in Fig.~\ref{fig:pt_diagram}).
To locate the critical point in the NS calculations, we drew on the results of Bruce and
Wilding~\cite{Bruce_Wilding} and calculated the density distribution in the
temperature region of the peak. With this, we estimate the critical parameters to be
$p_c=0.1-0.2~\mathrm{GPa}$ and $T_c=9500-10600~\mathrm{K}$.
The low-pressure melting transition is estimated to be $\approx 1650\mathrm{~K}$, hence
underestimating the experimentally determined transition.
This inaccuracy could be either due to our GAP or to an inherent error of the PBE
functional used to train it. Since NS calculations at the PBE level are simply intractable,
there is no straightforward way to pinpoint the origin of this disagreement with experiment. 
The NS calculations found the solid structure to be \gls{fcc} as expected,
and explored other close-packed stacking variants only in thermodynamically insignificant
proportions. 

\subsection{Spontaneous FCC nucleation and crystallization}\label{sec:spontaneous_nucleation}

\begin{figure}[t]
    \centering
    \includegraphics[width=\columnwidth]{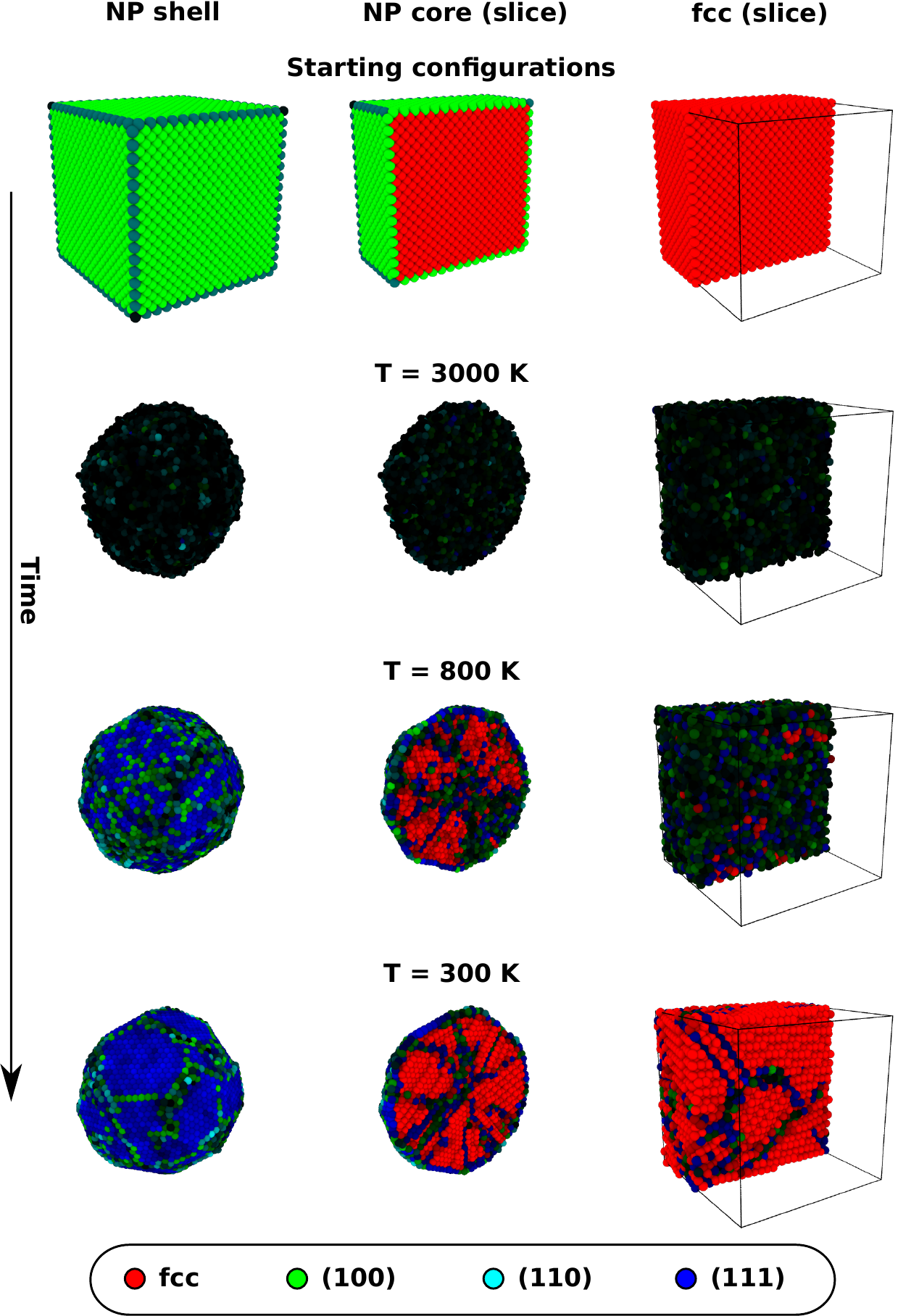}
    \caption{Snapshots throughout the process of spontaneous crystallization from a melted
    Pt droplet as it cools down to room temperature, as modeled with our GAP. The left column
    shows the resulting \gls{NP} from the outside, whereas the central column shows a slice through
    the middle. The same process for bulk Pt is shown on the right column. The color coding
    indicates the degree of similarity, computed from SOAP kernels,
    of each local atomic environment (centered on the
    atoms) to the stable bulk \gls{fcc} motif, as well as the three most common surface motifs: (100),
    (110) and (111), where (111) is the most stable facet. The dark bands
    between \gls{fcc} (red) regions in the final structures correspond to grain boundaries.}
    \label{fig:surface_nanoparticle}
\end{figure}

\begin{figure}[t]
    \centering
    \includegraphics[width=\columnwidth]{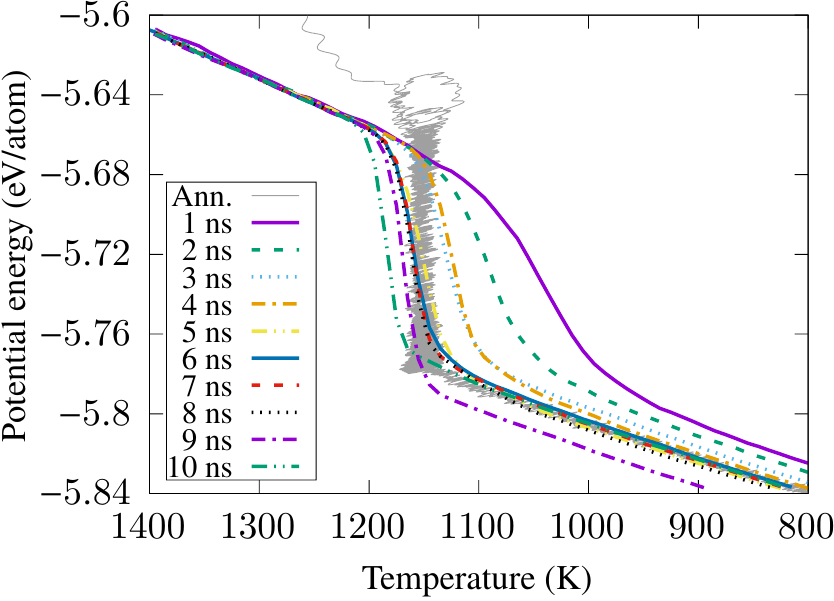}
    \caption{Potential energy profile as a function of temperature in a series of
    melt-quench simulations, for different cooling rates (1~ns to 10~ns cooling period). The overall
    process starts at 3000~K and ends at 300~K; the shown data focuses on the region
    where crystallization takes place, corresponding to the formation of stable \gls{fcc} motifs.
    The thin gray line shows the profile of a simulation where the sample is quenched
    extremely fast from 3000~K to 1150~K and annealed at that temperature before being brought
    down to room temperature. See text for details.}
    \label{fig:spontaneous_crystallization}
\end{figure}

We also used the Pt GAP to study the spontaneous nucleation of the stable \gls{fcc} structure and the
spontaneous formation of facets in a large \gls{NP} (16384 atoms) with MD.
Figure~\ref{fig:surface_nanoparticle}
shows the sequence from the initial cube carved out of an \gls{fcc} lattice. This is melted at
3000~K for 40~ps and then the quenching process takes place by cooling the \gls{NP} from 3000~K
down to 300~K over 1~ns using a linear temperature profile, controlled by a Berendsen
thermostat with time constant 0.1~ps. The figure also shows a slice through the middle of
the \gls{NP} and, for reference, a periodic solid with the same number of atoms and undergoing
the same temperature
profile. For the solid, the pressure is controlled with a Berendsen barostat with time
constant 1~ps and inverse compressibility equal to 100 times that of water.

To get further insight into the atomistic processes taking place during crystallization,
in Fig.~\ref{fig:surface_nanoparticle} we map the similarity of the local atomic structures
to reference atomic motifs: bulk \gls{fcc}
and the stable (100), (110) and (111) \gls{fcc} surface reconstructions. This is done by
computing the SOAP descriptors of each atom in the system and calculating the similarity
kernel with the SOAP descriptors of the reference motifs. These similarities are indicated
by color coding the resulting structures.
As expected, towards the end of the quench the interior of the \gls{NP} (as well as the solid) is
\gls{fcc}-like, and the \gls{NP} facets are (111)-like. Interestingly, the simulation shows that the
formation of the \gls{fcc} interior is nucleated from the surfaces inwards. Therefore, there is
grain formation with the (111) direction pointing approximately from the surface towards
the center of the \gls{NP}. For this reason, the resulting \gls{NP} is polycrystalline, with the grain
boundaries indicated by dark-colored atoms. It is clear from the figure that the formation
of the \gls{fcc} interior in the \gls{NP} happens at a \textit{higher} temperature than in the
solid due to the nucleation effect at the (111) facets. A video animation of this process
is available~\cite{caro_2022b}.

To elucidate the role of quench rate on the results, we monitored
the evolution of the \gls{NP}'s structure as it was cooled down from 3000~K to 300~K for
additional quench rates corresponding to 2~ns to 10~ns simulations, with the same MD settings as before.
Figure~\ref{fig:spontaneous_crystallization} shows the evolution of the potential energy as a
function of temperature in the 1400~K to 800~K temperature window, where most of the \gls{fcc} nucleation
takes place in these simulations (outside of this range the potential energy evolves linearly
with temperature, as expected
from the virial theorem). According to our MD results, the onset of significant structure rearrangement
favorable towards \gls{fcc} nucleation takes place at around 1200~K and continues down to a temperature which
depends on the quench rate (the slower the rate the higher the final temperature). From these values we
infer an optimal crystallization temperature around 1150~K. This is analogous to the graphitization
temperature in carbon materials~\cite{detomas_2019,wang_2022}. We therefore repeated the MD simulation
starting from the 3000~K melted \gls{NP} but fixing the thermostat's target temperature at 1150~K
and annealed for 1~ns (indicated as ``Ann.'' in the figure). There is a rapid quench from 3000~K
to 1150~K and then the system equilibrates for a few ps, corresponding to the loop seen at high
potential energy, before it starts to go down in energy as it crystallizes (the vertical drop in
potential energy at 1150~K). Most of the annealing process was completed after
250~ps, with no noticeable
further drop in potential energy after 500~ps of MD. After the 1~ns annealing simulation had ended,
we further quenched the structure to 300~K over 100~ps using a linear temperature profile. The
results showed good agreement with the more computationally demanding slow quenches. This
annealing process at 1150~K thus allows us to minimize the number of MD steps that are required to
generate a reasonably stable \gls{NP}, generated from a process mimicking spontaneous solidification.


\section{Conclusions and outlook}\label{sec:conclusions}
We have developed a \gls{GAP} for Pt with state-of-the-art force-field accuracy for the description of
bulk, surface and nanostructured systems. We have benchmarked our \gls{GAP} against
PBE-DFT for general accuracy, elasticity, phonons, surface energetics and \gls{NP} formation
energies. Except for small \glspl{NP} ($N_\text{atoms} \lesssim 40$), our \gls{GAP} shows
remarkable agreement with the reference PBE-DFT data. We have then proceeded to use the \gls{GAP} in
situations beyond the reach of PBE-DFT calculations. Namely, we have computed the
temperature-pressure phase diagram and studied the spontaneous solidification and \gls{fcc}-motif nucleation
in a large \gls{NP}. The new \gls{GAP} and several other resources have been made freely available.
In the near future, we will further develop our reference database and
the potential itself for improved description of \glspl{NP} and surface dynamics, with the
objective to get detailed insight into the atomic-scale phenomena taking place in Pt-based systems
of interest in (electro)catalysis.


\begin{acknowledgments}\label{sec:acknowledgements}
J.~K. and M.~A.~C. gratefully acknowledge funding from the Academy of Finland under
the C1 Value Programme, project No. 329483. M.~A.~C. also acknowledges personal
funding from the Academy of Finland, project No. 330488. 
H.~J. acknowledges funding from the Icelandic Research Fund, project No. 207283-053.
L.~B.~P. acknowledges support from the EPSRC through an Early Career Fellowship (EP/T000163/1).
Computational resources
for this project were obtained from CSC - IT Center for Science and Aalto University's
Science-IT project. 
\end{acknowledgments}

\appendix

\section{VASP input file}

The VASP INCAR input file used for the PBE-DFT calculations is given below:
\begin{verbatim}
PREC = Accurate
ENCUT = 520
EDIFF = 1.0e-05
ISMEAR = 0; SIGMA = 0.1
ALGO = Normal
LWAVE = .FALSE.
LCHARG = .FALSE.
\end{verbatim}
The $k$-space sampling is not explicitly set in the INCAR file. Instead, \textbf{k}
points are chosen by homogeneously sampling the first Brillouin zone
with the total number of points determined by the relation
$n_\text{atoms} \times n_\textbf{k} = 1000$. To enable high-throughput calculations,
the Fireworks framework~\cite{jain_2015} was used
for task automation and single-point workflows, similar to the implementation in Atomate~\cite{mathew_2017},
which rely on Custodian~\cite{ong_2013} as VASP handler.

\end{document}